\documentclass{moriond}
\usepackage{amsmath}
\usepackage{microtype}

\def\Journal#1#2#3#4{{#1} {\bf #2}, #3 (#4)}

\def\NPB{{\em Nucl. Phys.} B}
\def\PLB{{\em Phys. Lett.}  B}
\def\PRL{\em Phys. Rev. Lett.}
\def\PRD{{\em Phys. Rev.} D}

\def\be{\begin{equation}}
\def\ee{\end{equation}}
\def\bea{\begin{eqnarray}}
\def\eea{\end{eqnarray}}

\newcommand{\wbar}{\overline{w}}

\newcommand{\Photo}{}

\begin{document}
\vspace*{3.5cm}
\title{ON MULTIPLE GLUON EXCHANGE WEBS}

\author{ M. HARLEY }

\address{Higgs Centre for Theoretical Physics, University of Edinburgh,\\
Edinburgh EH9 3FD, Scotland}
\!
\maketitle\abstracts{
	I present an overview of the study of infrared singularities through the eikonal approximation and the concept of webs. 
	 Our work reveals the interesting structure of an infinite subclass of webs, Multiple Gluon Exchange Webs. We find that they can be expressed as sums of products of functions depending upon only a single cusp angle, spanned by a simple basis of functions, and conjecture that this structure will hold to all orders.}

\!
\vspace{-0.5cm}
\section{Introduction}
The scattering amplitudes of gauge theories suffer from so called infrared (IR) singularities where massless particles carry vanishing momenta. However, observables obtained from such theories are rendered finite by an intricate cancellation of IR singularities coming from real and virtual contributions order-by-order in the perturbative series. It is therefore necessary to compute these singularities in order to develop subtraction algorithms for Monte Carlo phase-space integration and for the resummation of large-logarithmic corrections to particular collider observables. The study of IR singularities also has interesting implications for the theoretical study of non-abelian gauge theories. Being simpler to calculate than complete scattering amplitudes, their study promises to further our understanding of the mathematical structure of scattering amplitudes deep into their perturbative expansion.

At present, for general kinematics, the IR singularities of QCD are known up to three loop order for two hard partons \cite{cusp3} and up to two loops for multiparton scattering \cite{ads}. Our goal is to push this forward to a full determination of the three loop soft anomalous dimension. In the following, we shall be focusing on the virtual IR singularities of QCD amplitudes, where all hard partons are massive. This solves a more general problem than a massless treatment, for example the singularities of amplitudes containing top quarks, while allowing the IR singularities of amplitudes containing hard massless particles to be reproduced by taking the lightlike limit of the relevant partons. When all are taken to the lightlike limit, this will allow us to determine corrections to the sum-over-dipoles formula for multiparton scattering of massless hard partons at three loops \cite{dip}.

\!
\section{Eikonal Approximation and Factorisation}

 In order to study IR singularities at high loop-order without the overburdening complications inherent to full amplitudes we make some approximations. We note that a virtual gluon carrying no momentum can not possibly affect the spin or momentum of the emitting hard partons, allowing us to discard such information from our calculation. In doing so we find that the IR singularities of massive scattering amplitudes are mapped one-to-one with the singularities of Wilson line correlation functions,
\begin{eqnarray}
	S = \langle \Phi_{\beta_1}(0,\infty) \otimes \Phi_{\beta_2}(0,\infty) \otimes \ldots \otimes \Phi_{\beta_n}(0,\infty) \rangle\, .
\end{eqnarray}
Here the scaleless velocities, $\beta_i$, are defined by $p_i = Q \beta_i$ for each hard parton momentum $p_i$ and for an arbitrary scale $Q$. A Wilson line is defined by,
\begin{eqnarray}
	\Phi_{\beta_i}(a, b) = \mathcal{P} \exp\bigg({\rm i}g_s \int_a^b ds\, \beta_i \cdot A(s\beta_i)\bigg)\, ,
\end{eqnarray}
where, $\mathcal{P}$ indicates ordering of colour matrices along the path and $A_\mu = A^a_\mu T^a_i$ is the gauge field in the representation of the parton $i$. This can be thought of as replacing the hard, coloured partons with classical, straight-line, semi-infinite radiators of soft gluons, carrying only a direction and colour information. This is known as the eikonal approximation. Interestingly, in discarding scale from the problem we will find that even in the renormalized QCD theory, we have introduced new ultraviolet (UV) poles at the cusps where Wilson lines meet. We are thus required to renormalize the soft function,
\begin{eqnarray}
	S_{\text{ren.}}(\alpha_{ij}, \alpha_s(\mu_R^2), \mu, m) = Z(\alpha_{ij}, \alpha_s(\mu_R^2), \epsilon, \mu) S(\alpha_{ij}, \alpha_s(\mu_R^2), \epsilon, m)\, ,
	\label{eq:multrenReg}
\end{eqnarray}
where $\alpha_{ij}$ are related to the cusp angles through,
\begin{equation}
	\frac{2\beta_i\cdot \beta_j}{\sqrt{\beta_i^2 \beta_j^2}} = -\alpha_{ij} - \frac{1}{\alpha_{ij}}\, ,
\end{equation}
$\alpha_s(\mu_R^2)$ is the renormalised strong coupling, $\epsilon$ is the dimensional regularisation parameter and $\mu$ is the renormalisation scale. To calculate the UV singularities of $S$ it is necessary to introduce some infrared regulator, $m$, in order to disentangle the UV and IR poles beyond one loop. When an IR regulator is not present, however, all loop diagrams contributing to $S$ will be scaleless Feynman integrals in dimensional regularisation and therefore zero. This implies that a cancellation must be occurring between the UV and IR divergences, hence in determining $Z$ we also obtain the IR poles of $S_{\text{ren.}}$,
\begin{eqnarray}
	S_{\text{ren.}}(\alpha_{ij}, \alpha_s(\mu_R^2), \mu) = Z(\alpha_{ij}, \alpha_s(\mu_R^2), \epsilon, \mu)\, .
\end{eqnarray}
Therefore, we are able to express the amplitude in a factorized form,
\begin{eqnarray}
	\mathcal{M}(p_i/\mu, \alpha_s(\mu_R^2), \epsilon) = Z(\alpha_{ij}, \alpha_s(\mu_R^2), \mu, \epsilon)~ H(p_i/\mu, \alpha_s(\mu_R^2)) \, ,
\end{eqnarray}
where $H$ is known as the hard function, a matching coefficient which can be determined at each order by computing the full amplitude, and all of the infrared singularities are contained within the renormalisation factor.

\!
\section{Non-abelian Exponentiation in Multiparton Scattering}
\label{sec:NonAbExp}
A renormalisation group equation for the soft function can be deduced from Eq.~(\ref{eq:multrenReg}),
\begin{eqnarray}
	\frac{d}{d\ln \mu^2} S_{\text{ren.}} =  - \Gamma S_{\text{ren.}}\, ,
	\label{eq:SrenEqn}
\end{eqnarray}
defining $\Gamma$, the well known `soft anomalous dimension'. Eq.~(\ref{eq:SrenEqn}). The solution of this equation implies exponentiation of $S_\text{ren.}$. There is an independent interpretation of exponentiation  which is entirely diagrammatic as was first realised in the context of QED \cite{yen}, generalised to the non-Abelian case in the context of Wilson loops \cite{gft}, and in then later generalised further to the multi-leg case \cite{rep}. In the case of the cusp (two Wilson lines), the exponent is simply the sum of all irreducible diagrams, in the sense that they cannot be separated into subdiagrams by cutting Wilson lines alone. Such diagrams are known as webs. In multiparton scattering, however, the situation is more complicated owing to non-trivial colour flow at the hard interaction vertex. Here, the notion of webs is generalised to encompass sets of (possibly reducible) diagrams related by permutation of ordering of gluon emissions along the lines.

Webs in both the cusp and multiparton processes appear in the exponent with an `exponentiated colour factor' (ECF), rather than the conventional colour factors of each diagram. In the multiline case a web, $W$, can be expressed according to,
\begin{eqnarray}
	W = \sum_{D,D'} C(D) R_{D D'} \mathcal{F}(D')
\label{eq:defWeb}
\end{eqnarray}
where $D$, $D'$ index the diagrams in the web, $C(D)$ and $\mathcal{F}(D)$ are the diagram colour and kinematic factors, respectively, and $R_{D D'}$ is a mixing matrix which can be determined using the methods produced in \cite{rep,gsw2}. It has been proven that ECFs are always given by colour factors of fully connected diagrams \cite{gsw2}.



\!
\section{Subtracted webs}
As mentioned above, we can determine the IR singularities of the soft function by renormalising the Wilson line correlator. Considering the multiplicative renormalisation of the soft function in eq.~(\ref{eq:multrenReg}), the renormalisation factor, $Z$, can also be expressed as an exponential of counterterms. As $Z$ and $S$ are both matrix valued in multiparton scattering, we are required to use the Baker-Campbell-Hausdorff formula to find the exponent of $S_\text{ren.}$. Along with the physical stipulation that the anomalous dimension itself is finite, this allows us to express $\Gamma$ in terms of the coefficients of the Laurent series of the webs at each order in $\alpha_s$,
\be
	w = \sum_{n=1}^\infty w^{(n)} \alpha_s^n = \sum_{n=1}^\infty \sum_{k=-n}^\infty w^{(n,k)} \alpha_s^n \epsilon^k\, .
\ee
We can then write \cite{gsw} the perturbative expansion of the anomalous dimension $\Gamma = -2 \sum_{n=1}^\infty n \wbar^{(n)} \alpha_s^n\, ,$
where we define `subtracted webs',
\begin{eqnarray}
	\wbar^{(1)} &=& w^{(1,-1)}\, , \\
	\wbar^{(2)} &=& w^{(2,-1)} + \frac12 \big[w^{(1,-1)}, w^{(1,0)} \big]\, , \\
	\wbar^{(3)} &=& w^{(3,-1)} - \frac12 \big[w^{(1,0)}, w^{(2,-1)} \big] - \frac12 \big[w^{(2,0)}, w^{(1,-1)} \big] \nonumber \\
	& & - \frac16 \big[w^{(1,0)}, \big[w^{(1,-1)}, w^{(1,0)} \big] \big] + \frac16 \big[w^{(1,-1)}, \big[w^{(1,-1)}, w^{(1,1)} \big] \big]\, . \\
	\ldots & & \nonumber
\end{eqnarray}
The contents of the commutators in each subtracted web are composed of the subdiagrams of the web $w^{(n,-1)}$. Subtracted webs have several useful properties lacking in webs such as independence of the IR regulator and a corresponding restoration of crossing symmetry, which is broken by IR regularisation.

\!
\section{Multiple Gluon Exchange Webs}
After having discussed webs in general we now wish to restrict ourselves to the simplest class of webs contributing to the multiparton soft anomalous dimension. Multiple Gluon Exchange Webs (MGEWs) comprise only webs which have gluons exchanged directly between Wilson lines, lacking gluon self-interaction away from the Wilson lines.
Since we are interested in the soft anomalous dimension we shall consider subtracted MGEWs, which in general \cite{mgew} have the structure,
\begin{eqnarray}
	\wbar^{(n)} = \left(\prod_{i=1}^n r(\alpha_i)\right) \sum_{j} c_j^{(n)} \int_0^1 \left[\prod_{k=1}^n dx_k \left(\frac{1}{x_k - \frac{1}{1-\alpha_k}} - \frac{1}{x_k+\frac{\alpha_k}{1-\alpha_k}} \right) \right] \mathcal{G}_{j}(\{x_i\}) \Theta_j(\{x_i\})\, ,
\end{eqnarray}
where $c_j^{(n)}$ the connected ECFs described in section \ref{sec:NonAbExp}, $\mathcal{G}_j$ are the subtracted web kernels corresponding to each of these colour factors, $r(\alpha) = (1+\alpha^2)/(1-\alpha^2)$ and $\Theta_j(\{x_i\})$ are the residual Heaviside functions left from the ordering of gluons on the Wilson lines. The subtracted web kernel can be obtained algorithmically following the methodology outlined in \cite{mgew}.

Based on physical constraints on the analytic structure of subtracted webs it has been conjectured \cite{eg} that subtracted MGEWs should be expressible as sums of products of functions of a single angle. Thus giving them a structure remarkably similar to that found in the cusp anomalous dimension \cite{hh}. Though generally the diagrams contributing to a subtracted MGEW contain multiple-polylogarithms, $\mathcal{G}_j$ is composed of only sums of products of logarithms in every subtracted web thus far considered \cite{mgew}, lending further evidence to this conjecture. Moreover, we find that all subtracted MGEWs can be expressed in terms of a remarkably simple basis of functions,
\begin{equation}
	M_{k,l,m}(\alpha) = \frac{1}{r(\alpha)} \int_0^1 dx \left(\frac{1}{x - \frac{1}{1-\alpha}} - \frac{1}{x+\frac{\alpha}{1-\alpha}} \right) \ln^k\left(\frac{q(x,\alpha)}{x^2}\right) \ln^l\left(\frac{x}{1-x}\right) \ln^m \widetilde{q}(x,\alpha)\, ,
\label{eq:basis}
\end{equation}
where,
\begin{eqnarray}
	\ln q(x,\alpha) = \ln\big( 1 - (1-\alpha)x \big) + \ln\left(1 - \left(1-\frac{1}{\alpha}\right)x\right)\, , \\
	\ln \widetilde{q}(x,\alpha) = \ln\big( 1 - (1-\alpha)x \big) - \ln\left(1 - \left(1-\frac{1}{\alpha}\right)x\right)\, .
\end{eqnarray}
which we have confirmed to hold in all subtracted MGEWs up to three loops, and even in the case of two four loop examples. We conjecture that this basis will fit all subtracted MGEWs \cite{mgew}. It still remains to fully understand the simple analytic structure of subtracted MGEWs and prove the above conjectures. We are also continuing to take strides towards an understanding of webs in general, and are progressing towards a calculation of the multiparton three loop soft anomalous dimension in general kinematics. More generally, this research goes beyond the computation of higher order corrections and aims at exploring the all-order structure of IR singularities in gauge theories and the associated physics.
\!
\section*{Acknowledgments}
\vspace{-0.2cm}
Many thanks go to my collaborators Giulio Falcioni, Einan Gardi, Lorenzo Magnea and Chris White. This work was supported by The University of Edinburgh.

\end{document}